\documentclass[12pt]{article}

\usepackage{epsfig}
\usepackage{ams math}
\newcommand{\ft}[2]{{\textstyle\frac{#1}{#2}}}

\newcommand{\eqn}[1]{(\ref{#1})}

\font\myb=msbm10 at 12pt
\def\b#1{\hbox{\myb#1}}

\newif\ifpdf
\ifx\pdfoutput\undefined
\pdffalse 
\else
\pdfoutput=1 
\pdftrue
\fi

\begin{document}
\begin{titlepage}
\begin{center}
\hfill ITP-UU-02/02  \\
\hfill SPIN-02/04  \\
\hfill YITP-SB-02-10  \\
\hfill {\tt hep-th/0202149}\\
\vskip 10mm

{\Large {\bf Hyperk\"ahler Cones \\[3mm]
and Orthogonal Wolf Spaces }}

\vskip 8mm

{\bf Lilia Anguelova$^{*}$, Martin Ro\v{c}ek$^{*}$  and
Stefan Vandoren$^{\dag}$ }

\vskip 3mm

$^*${\em C.N. Yang Institute for Theoretical Physics}\\
{\em SUNY, Stony Brook, NY 11794-3840, USA}\\[2mm]
{\tt anguelov@grad.physics.sunysb.edu}\\
{\tt Ro\v cek@insti.physics.sunysb.edu}\\[2mm]
$^\dag${\em Institute for Theoretical Physics} and {\em Spinoza Institute}\\
{\em Utrecht University, Utrecht, The Netherlands}\\
{\tt vandoren@phys.uu.nl}\\

\vskip 5mm

\end{center}

\vskip 2mm

\begin{center} {\bf ABSTRACT } \end{center}
\begin{quotation}
\noindent{We construct the hyperk\"ahler cones corresponding to the quaternion-K\"ahler orthogonal Wolf spaces $\frac{SO(n+4)} {SO(n)\times SO(4)}$ and their non-compact versions, which appear in hypermultiplet couplings to $N=2$ supergravity.  The geometry is completely encoded by a single function, the hyperk\"ahler potential, which we compute from an $SU(2)$ hyperk\"ahler quotient of flat space.  We derive the Killing vectors and moment maps for the $SO(n+4)$ isometry group on the hyperk\"ahler cone. For the non-compact case, the isometry group $SO(n,4)$ contains $n+2$ abelian isometries which can be used to find a dual description in terms of $n$ tensor multiplets and one double-tensor multiplet. Finally, using a representation of the hyperk\"ahler quotient via quiver diagrams, we deduce the existence of a new eight dimensional ALE space.}
\end{quotation}

\vfill
\flushleft{\today}
\end{titlepage}
\eject
\section{Introduction}
\setcounter{equation}{0}
Quaternion-K\"ahler (QK) manifolds have recently attracted a lot of attention
in the context of $N=2$ gauged supergravity, both in four and five spacetime 
dimensions. In particular, those with negative scalar curvature appear as target spaces for $N=2$ hypermultiplets coupled to supergravity 
\cite{BagWit}.  The {\it homogeneous} QK spaces $G/H$  are classified by Wolf and Alekseevskii \cite{WA}; the compact ones are given by the three infinite series
\begin{equation}
\b{H}{\rm{P}}(n)=\frac{Sp(n+1)}{Sp(n)\times Sp(1)}, \,\,\, X(n)=\frac{SU(n+2)}{SU(n)\times U(2)}, \,\,\, Y(n)=\frac{SO(n+4)}{SO(n)\times SO(4)} \ ,
\end{equation}
of dimension $4n$, and the five exceptional cases
\begin{equation}
\frac{G_2}{SO(4)}, \,\,\, \frac{F_4}{Sp(3)\times Sp(1)}, \,\,\, \frac{E_6}{SU(6)\times Sp(1)}, \,\,\, \frac{E_7}{Spin(12)\times Sp(1)}, \,\,\, \frac{E_8}{E_7\times Sp(1)}\ ,
\end{equation}
of dimensions 8, 28, 40, 64 and 112 respectively.  Their non-compact versions have negative scalar curvature and hence are relevant for supergravity applications. For low dimensions, there are relations between  these spaces: the four-sphere $Y(1)\cong \b{H}\rm{P}(1)$, and $Y(2)\cong X(2)$. These isomorphisms are discussed in more detail below. 

A complete classification of QK spaces does not exist. Because their Sp(1) 
curvature does not vanish, and their quaternionic two-forms are not closed, 
QK spaces are rather difficult to deal with. This complicates the study of 
hypermultiplet couplings to supergravity, which arise in the low energy effective action of type II superstring compactification on Calabi-Yau 
threefolds \cite{CFG}. In this paper, we focus on the $Y(n)$ spaces.
Their noncompact versions
\begin{equation}
\tilde Y(n)=\frac{SO(n,4)}{SO(n)\times SO(4)}\ , \label{S}
\end{equation}
describe the classical moduli spaces of type II Calabi-Yau 
compactifications down to four dimensions, either as dual quaternionic 
spaces from the {\bf c}-map \cite{CFG}, or from dualizing $n$ tensor 
multiplets and a double-tensor multiplet \cite{BGHL}. 

Vector multiplet couplings to $N=2$ supergravity are geometrically much 
simpler. The scalar fields of the vector multiplets parametrize a 
special K\"ahler geometry in d=4 \cite{SKG} and a very special real geometry 
in d=5 \cite{VSRG}. Both these geometries are completely determined in 
terms of a single function from which the target space metric, and other geometrical quantities, can easily be computed. It is desirable to have a similar description for QK manifolds, such that the whole geometry is encoded by a single function of the hypermultiplet scalars. Such a description was first given in the mathematics literature, where it was shown \cite{Swann}, that to every $4n$ dimensional QK manifold one can associate a $4(n+1)$ dimensional hyperk\"{a}hler manifold which admits a conformal homothety
and an isometric Sp(1)-action that rotates the three complex structures into each other. 
The conformal homothety $\chi^A$ satisfies 
\begin{equation}
D_{A}\chi^{B}=\delta_{A}^{B}\ , \qquad (A,B=1,...,4n)\ ,
\end{equation}
such that locally $\chi_A=\partial_A \chi$. The function $\chi$ is called the hyperk\"ahler potential and the metric is
\begin{equation}
g_{AB}=D_A\partial_B \chi\ .
\end{equation}
Such a manifold is a hyperk\"ahler cone (HKC) over a base which is an Sp(1) fibration over a QK space \cite{Swann, GR}.

It was later anticipated in \cite{Gal}, and shown explicitly in 
\cite{dWKV,dWRV1} how this connection is precisely realised in the conformal calculus for hypermultiplets. The transition from the HKC to the QK space is called the {\it $N=2$ superconformal quotient}, and is associated with eliminating the compensating hypermultiplet by an $SU(2)$ quotient and gauge-fixing combined with a reduction along the homothety. The simplest QK manifolds are the quaternionic projective spaces $\b{H}\rm{P}(n)$. Galicki has shown that the $X$ and $Y$ series can be obtained from these by performing quaternionic $U(1)$ and $SU(2)$ quotients, respectively \cite{G}.  It is interesting to study these spaces in terms of the corresponding HKC's. The HKC of $\b{H}{\rm{P}}(n-1)$ is simply $\b{H}^n \cong \b{C}^{2n}$, and the explicit construction of the HKC of $X(n-2)$ is a $U(1)$ hyperk\"ahler quotient of $\b{H}^n$ \cite{dWRV1}. 
In Section 2 of this paper, we derive the hyperk\"ahler potential
$\chi$ for the orthogonal Wolf spaces $Y(n-4)$ by performing an $SU(2)$ hyperk\"{a}hler quotient on $\b{H}^{n}$. The action of the $SU(2)$ on the coordinates of $\b{H}^n$ can be deduced from Galicki's quaternionic quotient \cite{G} because the diagram in Figure 1 is commutative \cite{Swann}. Hence quaternionic quotients between the QK manifolds lift to hyperk\"ahler quotients between the corresponding HKC's.
\begin{figure}[htbp]
\centering
\leavevmode
\epsfxsize=3.0in
\begin{center}
\leavevmode
\ifpdf
\epsffile{diagram1.pdf}  
\else
\epsffile{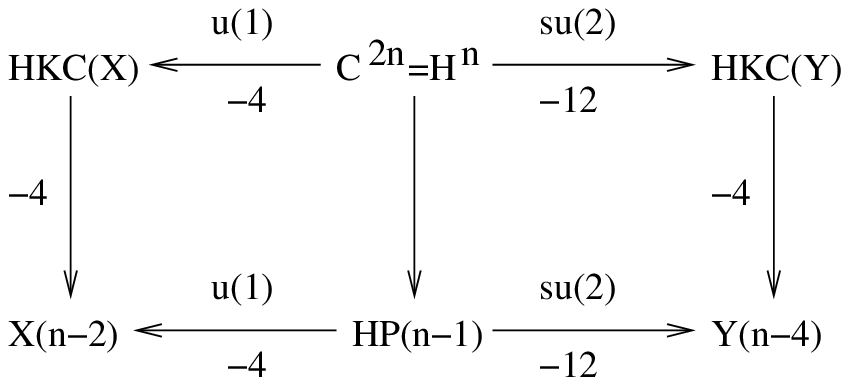}  
\fi
\end{center}
\caption{\footnotesize{The horizontal arrows stand for hyperk\"{a}hler 
quotients (upper), and quaternionic quotients (lower). The vertical arrows 
denote the superconformal quotients. The numerical label of each arrow denotes the change of the number of real dimensions accompanying the corresponding quotient.}}
\vspace{0.6cm}
\end{figure}
We also write down the moment maps and Killing vectors for the $SO(n)$ 
isometry group on the HKC. These appear in the scalar potential of gauged 
$N=2$ supergravity.

In Section 3 we consider the three special cases for which the hyperk\"{a}hler potential of the HKC of $Y(n-4)$ is known: $n=4$, where the HKC of $Y(0)$ is $\b{R}^{4}$, $n=5$, where the HKC of $Y(1)$ is $\b{R}^{8}$ and $n=6$, where the HKC of $Y(2)$ is the HKC of $X(2)$. We check our calculation by showing that each of these special cases agrees with the known results.

In Section 4 we construct the dual description of the HKC of the non-compact 
version $\tilde{Y}(n)$ in terms of $n$ tensor multiplets and a double tensor 
multiplet. This tensor multiplet description naturally appears in type IIB
compactifications.

In Section 5 we show how the hyperk\"ahler quotient can be lifted to $N=2$ superspace.  Once we have this description, we can use the techniques of 
\cite{LRvR} to write down the quiver diagrams for the HKC's of $Y(n)$. This point of view allows us to infer the existence of a new 8-dimensional ALE space.

Our long term goal is to understand the nonperturbative corrections to 
quaternionic geometries; we expect that these are most easily computed in terms of the hyperk\"ahler potential $\chi$.  We also note that HKC's appear as moduli spaces of Yang-Mills instantons \cite{Mac-B,V}, {\it e.g.}, the homogeneous QK $G/H$ manifolds determine the one-instanton moduli spaces with gauge group $G$ and stability group $H$. This suggests a connection between the moduli spaces of Calabi-Yau manifolds and Yang-Mills instantons \cite{V}.

\section{The SU(2) Quotient}
\setcounter{equation}{0}

In this section, we explicitly perform the non-abelian hyperk\"ahler quotient 
from $\b{H}^n=\b{R}^{4n}$ to the HKC of $Y(n-4)$, $n\geq 4$. For simplicity, we do the analysis for the compact case; the non-compact case is very similar and we give the final result at the end of this section. The space $\b{H}^{n}$ is flat, and has an isometry group $SO(n)$ which is linearly realized on the $n$ quaternions
\begin{equation}
q^M=\left( \begin{array}{cc} z_{+}^M & -\bar{z}_{-}^M \\ z_{-}^M & \bar{z}_{+}^M \end{array} \right)\ , \qquad  M=1,...,n\ .
\end{equation}
Clearly, $\b{H}^{n}$ is itself an HKC with hyperk\"ahler potential (or equivalently, $N=1$ superspace Lagrangian)
\begin{equation}
\chi\,(\b{H}^n)=\ft12{\rm Tr}\,\big(q^M{q^\dagger}^M)=z_+^M\bar z_+^M+z_-^M\bar z_-^M\ , \label{ch}
\end{equation}
where a sum over repeated indices is understood. The closed hyperk\"ahler two-forms on $\b{H}^n$ are
\begin{equation}
\Omega^3=-i\Big({\rm d}z_+^M\wedge {\rm d}\bar z_+^M+
{\rm d}z_-^M\wedge {\rm d}\bar z_-^M\Big)\ ,\quad
\Omega^+={\rm d}z_+^M\wedge{\rm d}z_-^M\ ,\quad \Omega^-=(\Omega^+)^*\ ,
\label{twoforms}
\end{equation}
such that the $SO(n)$ isometries $q^M\rightarrow O^M{}_N\,q^N$ are triholomorphic.

The infinitesimal $SU(2)$ action that we quotient by can be deduced from 
the quaternionic quotient  in \cite{G}; it acts by left multiplication on
each quaternion, with corresponding Killing vector fields on $\b{H}^n$
\begin{eqnarray}
k_{3}&=&iz_{+}^M\frac{\partial}{\partial z_{+}^M} - iz_{-}^M\frac{\partial}{\partial z_{-}^M} - i \bar{z}_{+}^M\frac{\partial}{\partial \bar{z}_{+}^M} +i\bar{z}_{-}^M\frac{\partial}{\partial \bar{z}_{-}^M}\ , \nonumber \\
k_{+}&=&-z_{-}^M\frac{\partial}{\partial z_{+}^M} + \bar{z}_{+}^M\frac{\partial}{\partial \bar{z}_{-}^M}\ ,\nonumber\\
k_{-}&=&z_{+}^M\frac{\partial}{\partial z_{-}^M} - \bar{z}_{-}^M\frac{\partial}{\partial \bar{z}_{+}^M}\ .
\end{eqnarray}
Hence, for each $M$, $(z_{+},z_{-})$ transforms as an $SU(2)$ 
doublet\footnote{Ordinarily, in a hypermultiplet the $N=1$ chiral 
superfields $z_+$ and $z_-$ transform in conjugate representations 
of the gauge group; however, since the spinor representation of $SU(2)$ is 
pseudoreal, one may consider an action that mixes them \cite{BS}. In the 
literature, this is sometimes called a ``half-hypermultiplet''.  In section 5 we discuss a formulation in terms of full hypermultiplets that lifts to an $N=2$ superspace description.}:
\begin{eqnarray}
\delta z_\pm\equiv \lambda^i k_i z_\pm~,\qquad\qquad
&\qquad&~~~\, i=3,+,- \nonumber\\
\delta z_{+} =i \lambda^3 z_{+} - \lambda^{+} z_{-} ~ ,~~  &\qquad& 
\delta \bar{z}_{+}=-i\lambda^3 \bar{z}_{+} - \lambda^{-} \bar{z}_{-}~,
\nonumber\\
\delta z_{-} = -i \lambda^3 z_{-} + \lambda^{-} z_{+} ~, &\qquad&
\delta \bar{z}_{-} = i \lambda^3 \bar{z}_{-} + \lambda^{+} \bar{z}_{+}~.
\label{var}
\end{eqnarray}
This isometry group is also triholomorphic\footnote{This $SU(2)$ should not be confused with the $Sp(1)=SU(2)$ that rotates the complex structure and is used to perform the $N=2$ superconformal quotient. Under its action $(z_+,\bar z_-)$ transforms as a doublet, which corresponds to right multiplication on each quaternion.}  and commutes with $SO(n)$.

The components of the corresponding triplets of moment maps\footnote{We follow the conventions and notations of \cite{dWRV1,dWRV2}; for each triholomorphic Killing vector field $k_I$, there is a triplet of moment maps $\vec{\mu}_I=(\mu^3_I,\mu_I^+,\mu_I^-)$.} on $\b{H}^n$ are, for $SU(2)$,
\begin{eqnarray}
\mu^3_i&=&-\Big(z_+\cdot \bar z_+-z_-\cdot \bar z_-,\,i\,z_-\cdot \bar z_+,
-i\,z_+\cdot \bar z_-\Big)\ ,\nonumber\\
\mu^+_i&=&\Big(-iz_+\cdot z_-,\,\ft12z_-\cdot z_-,\,\ft12 z_+\cdot z_+\Big)\ ,
\qquad \mu^-_i=(\mu^+_i)^*\ ,\label{SU2MM}
\end{eqnarray}
and for $SO(n)$ (with antihermitian generators $T_I$)
\begin{equation}
\mu_I^3=-i\Big((z_+T_I\bar z_+)+(z_-T_I\bar z_-)\Big)\ ,\quad
\mu_I^+=(z_+T_Iz_-)\ ,\quad \mu_I^-=(\mu_I^+)^*\ .\label{so-mm}
\end{equation}
As indicated by these formulae, we sometimes suppress indices in inner products and bilinear forms on $\b{H}^{n}$ to simplify the
notation. 

The $SU(2)$ hyperk\"ahler quotient requires setting the moment maps \eqn{SU2MM} to zero. Hence the independent set of constraints are the non-holomorphic ones
\begin{equation}
\bar{z}_{+} \cdot z_{+} = \bar{z}_{-} \cdot z_{-}\ ,\qquad 
\bar{z}_{+} \cdot z_{-} = 0\ , \qquad 
\bar{z}_{-} \cdot z_{+} = 0\ , \label{nonh}
\end{equation}
and the holomorphic ones 
\begin{equation}
z_{+} \cdot z_{-} = 0\ , \qquad  z_{+} \cdot z_{+} = 0\ ,\qquad z_{-} \cdot z_{-} = 0\ . \label{hol}
\end{equation}
To perform the quotient one has to solve (\ref{nonh}) and (\ref{hol}), reducing the number of real coordinates by $3+6=9$, and choose a gauge to fix the $SU(2)$ gauge freedom, further reducing the number of real coordinates by $3$; thus the dimension is reduced by $12$ real or $6$ complex coordinates, consistent with the dimension of the HKC of $Y(n-4)$, namely, $4(n-3)$. In practice it is hard to solve all the constraints; as explained in \cite{HKLR}, one may complexify\footnote{The complexification of the gauge group means doubling of the number of generators (to the set of initial Killing vectors ${k_{I}}$ one adds the set ${J^3 k_{I}}$, where $J^3$ is the complex structure determined by $\Omega^{3}(X,Y)=g(J^3X,Y)$).} the gauge group (in our case $SU(2)\rightarrow SL(2,\b{C}))$ and solve just the holomorphic constraints, which are preserved by the complexified gauge group. This construction is completely natural in $N=1$ superspace, where the non-holomorphic constraints are solved for the $N=1$ gauge fields. If we choose 3 {\it complex} gauge conditions and impose the holomorphic constraints, we see that we still reduce the dimension by 12. The hyperk\"ahler potential on the quotient space is found by gauging the $N=1$ Lagrangian for $\b{H}^n$ and substituting the solution for the gauge fields.

Recall that the hyperk\"ahler potential for the flat space $\b{H}^n$ is (\ref{ch}). For each value of the index $M$ the pair $(z_+^M, z_-^M)$ is a doublet under the $SU(2)$ which we quotient by (see (\ref{var})). Hence we have $n$ copies of the $2$-dimensional represenation of $SU(2)$ in (\ref{ch}). 

We introduce a gauge field ${\rm e}^V$, such that the gauged hyperk\"ahler potential of $\b{H}^n$ is (\cite{BS,LR,HKLR}, see also section five and appendix B of \cite{dWRV1}):
\begin{equation}
\hat \chi=\sum_{M=1}^n \left( \bar{z}_{+}^M \,\, \bar{z}_{-}^M \right) 
\Big( \begin{array}{cc} {\rm e}^V \end{array} \Big)
\left( \begin{array}{c} 
z_{+}^M \\ z_{-}^M \end{array} \right)\ , \label{hatchi}
\end{equation}
where $\Big( \begin{array}{cc} {\rm e}^V \end{array} \Big)$ is the $2\times 2$ matrix of $N=1$ $SU(2)$ gauge fields and hence is hermitian with determinant $1$. The nonholomorphic constraints (\ref{nonh}) ({\it c.f.} (\ref{SU2MM})) can be written as
\begin{equation}
\mu_i^3=\left( \bar{z}_{+}^M \,\, \bar{z}_{-}^M \right) T_{i} 
\left( \begin{array}{c} z_{+}^M \\ z_{-}^M \end{array} \right) = 0\ , \qquad
i=3,+,-
\end{equation}
where $T_{i}$ are the $SU(2)$ generators normalized as
\begin{equation}
T_{3}=\left( \begin{array}{cc} 1 & 0 \\ 0 & -1 \end{array} \right)\ , 
\qquad T_{+}=\left( \begin{array}{cc} 0 & 1 \\ 0 & 0 \end{array} 
\right)\ , 
\qquad T_{-}=\left( \begin{array}{cc} 0 & 0 \\ 1 & 0 \end{array} 
\right)\ . \label{gen}
\end{equation}
The gauged nonholomorphic moment maps are \cite{BS,LR,HKLR}:
\begin{equation}
\hat \mu_i^3=\left( \bar{z}_{+}^M \,\, \bar{z}_{-}^M \right) 
T_{i}\, \Big( \begin{array}{cc} {\rm e}^V \end{array} \Big)
\left( \begin{array}{c} z_{+}^M \\ z_{-}^M \end{array} \right) = 0\ . 
\label{mug}
\end{equation}
Solving (\ref{mug}) for the gauge fields we find
\begin{equation}
\Big( \begin{array}{cc} {\rm e}^V \end{array} \Big) =\frac1{\sqrt{(\bar{z}_{+}\cdot z_{+})(\bar{z}_{-}
\cdot z_{-}) - (\bar{z}_{+}\cdot z_{-})( \bar{z}_{-}\cdot z_{+})}}
\left( \begin{array}{cc} \bar{z}_{-}\cdot z_{-} & -\bar{z}_{-}\cdot z_{+}\\ 
-\bar{z}_{+}\cdot z_{-} & \bar{z}_{+}\cdot z_{+}\end{array} \right)\ , 
\end{equation}
and hence the hyperk\"{a}hler potential of the quotient is
\begin{equation}
{\chi}_{quotient}=2 \sqrt{\bar{z}_{+}^M z_{+}^M \bar{z}_{-}^N z_{-}^N - 
\bar{z}_{+}^M z_{-}^M \bar{z}_{-}^N z_{+}^N}\ , \label{cq}
\end{equation}
with $M,N$ still running from $1$ to $n$. To find coordinates on the quotient we gauge-fix and solve the holomorphic constraints; we can choose
\begin{equation}
z_{+}^{n}=0\ , \qquad z_{-}^{n-1}=0\ , \qquad z_{+}^{n-2}=1\ . \label{gf}
\end{equation} 
Now the holomorphic constraints (\ref{hol}) can be easily solved:
\begin{equation}
z_{-}^{n-2}=-z_{+}^{a} z_{-}^{a}\ ,\qquad
z_{+}^{n-1}=i\sqrt{1+z_{+}^{a}z_{+}^{a}}\ , \qquad 
z_{-}^{n}=i\sqrt{z_{+}^{a}z_{-}^{a} z_{+}^{b} z_{-}^{b}+z_{-}^{a} z_{-}^{a}}\ ,
\label{MM-sol}
\end{equation}
where $\{z_+^a,z_-^b\},\,a,b=1,...,n-3$ are the coordinates on the HKC of $Y(n-4)$.  Thus the hyperk\"ahler potential on the HKC of $Y(n-4)$ is given by \eqn{cq}, subject to \eqn{gf} and \eqn{MM-sol}. Furthermore the dilatations act on $z_+$ and $z_-$ with scaling weights 0 and 2 respectively, and the holomorphic two-form on the HKC is simply
\begin{equation}
\Omega^+={\rm d}z_+^a\wedge {\rm d}z_-^a\ ,\qquad a=1,...,n-3\ .
\label{HKC-2form}
\end{equation}
The gauge choice \eqn{gf} is legitimate, but not always the most convenient, as we see in the next section.

The moment maps of the $SO(n)$ isometry on the HKC can easily be computed from the $SU(2)$ quotient of \eqn{so-mm}. The $SU(2)$ gauged $SO(n)$ moment maps take a similar form to \eqn{mug}, and eliminating the gauge fields  yields
\begin{eqnarray}
\mu_I^3&=&\frac{2i}{\chi}\Big((\bar z_-\cdot  z_-)(\bar z_+T_I z_+)-(\bar z_+\cdot z_-)(\bar z_-T_I z_+)+(+\leftrightarrow -)\Big)\ ,\nonumber\\
\mu^+_I&=&(z_+T_Iz_-)\ ,\qquad \mu^-_I=(\mu^+_I)^*\ ,\label{SO-MM}
\end{eqnarray}
subject to the constraints \eqn{gf} and \eqn{MM-sol}.

The $SO(n)$ Killing vectors on the quotient space can also be easily 
computed. On $\b{H}^n$, the Killing vectors are simply 
$k^M{}_{I\pm}=(T_I)^M{}_Nz_{\pm}^N$, but after the quotient we have to add 
local compensating $SU(2)$ transformations to preserve the gauge 
conditions \eqn{gf}. We find
\begin{eqnarray}\label{comp-su2}
k^M{}_{I+}&=&(T_I)^M{}_Nz_+^N+(T_I)^n{}_Nz_+^N\,\frac{z_-^{n-2}z_+^M-z_-^M}{z_-^n}
-(T_I)^{n-2}{}_Nz_+^Nz_+^M\ ,\\
k^M{}{I-} &=& (T_I)^M{}_Nz_-^N-(T_I)^n{}_N\, \frac{z_+^Nz_-^{n-2}z_-^M}{z_-^n}
-(T_I)^{n-1}{}_N\,\frac{z_-^Nz_+^M}{z_+^{n-1}}+(T_I)^{n-2}{}_Nz_+^N
z_-^M\ .\nonumber 
\end{eqnarray}
One can verify that they preserve \eqn{gf} as well as 
\eqn{MM-sol}. 

These Killing vectors and moment maps form the main ingredients of
the scalar potential on the HKC. Performing the superconformal quotient 
leads to the Killing vectors and moment maps on $Y(n-4)$, which can
be compared to those given in \cite{FGPT}.

To end this section, we discuss the non-compact case; the QK space is now
\begin{equation}
\tilde Y(n-4)\equiv \frac{SO(n-4,4)}{SO(n-4)\times SO(4)}\ .
\end{equation}
The above formulae are modified by putting in a pseudo-Riemannian metric in the inner products: one starts with a metric $\eta_{M\bar N}$ on $\b{H}^{n-4,4}$ of signature $\eta_{M\bar N}={\rm diag}(----+\cdots+)$. The metric then used in the products $z^a\eta_{a\bar b}\bar z^b$ on the HKC of $\tilde{Y}(n-4)$ has signature $\eta_{a\bar b}={\rm diag}(-+\cdots+)$, such that, in our conventions, the metric on the QK space is positive definite.\footnote{This differs by an overall sign from the conventions of \cite{dWRV1,dWRV2}.}

\section{The Special Cases}
\setcounter{equation}{0}

In this section, we give more details about the $SU(2)$ gauge fixing, and 
compare our results with the known cases for $n=4,5,6$. It is convenient 
to introduce the $SU(2)$ gauge invariant variables (for all $n$),
\begin{equation}
z^{MN}\equiv  z_{+}^{M}z_{-}^{N} - z_{+}^{N} z_{-}^{M}\ . \label{inv}
\end{equation}
The hyperk\"ahler potential on the quotient \eqn{cq} is then, for all $n$,
\begin{equation}
\chi_{quotient}={\sqrt {2\,z^{MN}\bar z^{MN}}}\ ,
\end{equation}
and is manifestly $SU(2)$ gauge invariant. The variables \eqn{inv} satisfy 
a Bianchi-like identity,
\begin{equation}
z^{MN}z^{PQ}+z^{MQ}z^{NP}+z^{MP}z^{QN}=0\ , \label{Bianchi}
\end{equation}
and the holomorphic constraints \eqn{hol} imply
\begin{equation}
z^{MN}z^{NP}=0 \ .\label{new-constr}
\end{equation}
We show below how these constraints can be solved for $n=4,5,6$.

\subsection{$n=4$}

In this case the quaternion-K\"ahler manifold is just 
one point, and so the hyperk\"ahler cone above it must be $\b{R}^{4}$. The gauge choice \eqn{gf} does not give the canonical quadratic hyperk\"ahler potential on $\b{C}^2$, so we look for a different gauge. For $n=4$, any antisymmetric matrix can be decomposed into its selfdual and anti-selfdual part. Notice then that \eqn{Bianchi} and \eqn{new-constr} are solved by
\begin{equation}
z^{MN}=\pm \epsilon^{MNPQ} z^{PQ}\ ,
\end{equation}
so that we can parametrize
\begin{equation}
z^{MN}=u^i (\bar \sigma)_{ij}^{MN}u^j\ .\label{SDcoord}
\end{equation}
Here, $u^1,u^2$ are two complex coordinates and $ (\bar \sigma)_{ij}^{MN}$
are the standard selfdual matrices\footnote{We use the conventions
$\bar \sigma^{MN}=\ft12(\bar \sigma^M\sigma^N-\bar \sigma^N\sigma^M)$
with $\sigma^M=(\vec \tau,i)$ and $\bar \sigma^M=(\vec \tau,-i)$.  The $(\vec \tau)$ are the three Pauli matrices and $i,j$ indices are raised and lowered with $\epsilon_{ij}$ such that $(\bar \sigma)_{ij}^{MN}$ is symmetric. We could also choose the anti-selfdual matrices $\sigma^{MN}=\ft12(\sigma^M\bar \sigma^N-\sigma^N\bar \sigma^M)$, but this leads to equivalent results.}.

Using sigma-matrix algebra, it is easy to check that the constraints 
\eqn{Bianchi} and \eqn{new-constr} are satisfied and the hyperk\"ahler potential becomes simply 
\begin{equation}
\chi=4\sum_{i=1}^2 u^i\bar u^i\ ,
\end{equation}
which is the K\"ahler potential of $\b{C}^{2}$ with flat metric.

The gauge choice \eqn{SDcoord} for the bilinears $z^{MN}$ implies
\begin{equation}
z_{+}=\left(\, u,\,\, v,\,\, iu,\,\, iv\, \right)\ , \qquad 
z_{-}=\left(\, iv,\,\, -iu,\,\, v,\,\, -u \, \right). \label{z}
\end{equation}
with $u^1=u$ and $u^2=v$. Notice that this $SU(2)$ gauge solves 
{\it all} the constraints \eqn{nonh} and \eqn{hol}.

\subsection{$n=5$}
 
It is well-known that 
\begin{equation}
Y(1) = \frac{SO(5)}{SO(4)}= \frac{Sp(2)}{Sp(1) \times Sp(1)}=\b{H}\rm{P}(1)\ ,
\end{equation}
and hence the HKC above it must be $\b{H}^{2}=\b{R}^{8}$. The gauge \eqn{gf} again does not give the canonical, quadratic hyperk\"ahler potential on $\b{C}^4$, so we search for a more suitable gauge. 

The coordinates $z_{\pm}^{M}$ transform in the vector representation of $SO(5)$:
\begin{equation}
z_{\pm}^{M} \rightarrow  O^M{}_N z_{\pm}^{N}, 
\qquad M,N=1,...,5 \, , \label{so5}
\end{equation}
We rewrite this in $Sp(2)$ notation by using the Clifford algebra in four Euclidean dimensions, where all gamma matrices can be taken hermitian. We define $4\times4$ matrices
\begin{equation}
(v_{\pm})_{ij}\equiv -\ft12 z_{\pm}^M{\gamma^M}_i{}^k {\cal C}_{kj}
\qquad i,j=1,\ldots,4\ ,\label{def-v}
\end{equation}
where $\gamma^5=\gamma^1\gamma^2\gamma^3\gamma^4$ and 
${\cal C}$ is the charge conjugation matrix, which can be taken antisymmetric and with square $-1$, such that $(\gamma^M {\cal C})_{ij}$ is antisymmetric.  ${\cal C}$ serves as an $Sp(2)$ metric and can be used to lower and raise $(i,j)$ indices.  Hence the matrices $v_{\pm}$ are antisymmetric; the relation $({\cal C}^{-1})^{ij}v_{\pm ij}=0$, then leaves five independent components.  We can invert \eqn{def-v},
\begin{equation}
z_{\pm}^M=-\ft12 {\rm Tr} \Big( \gamma^M v_{\pm} {\cal C}^{-1} \Big)\ . 
\label{invert}
\end{equation}
Now we introduce $SU(2)$ gauge invariant variables analogous to (\ref{inv}):
\begin{equation}
w_i{}^j\equiv \Big(v_{+} {\cal C}^{-1} v_{-} {\cal C}^{-1} - 
v_{-} {\cal C}^{-1} v_{+} {\cal C}^{-1}\Big)\ .
\end{equation}
The matrix $w_{ij}\equiv (w\,{\cal C})_{ij}$ is symmetric; the relation to 
\eqn{inv} is
\begin{equation}
w=\ft14 z^{MN}\gamma^{MN}\ .
\end{equation}
It is straightforward to show that the hyperk\"ahler potential \eqn{cq} becomes
\begin{equation}
\chi = \sqrt{ \, {\rm Tr} (w w^{\dagger})}\ ,\label{v-vdagger}
\end{equation}
up to an irrelevant overall factor.  In these variables, the moment map constraints (\ref{hol}) are
\begin{equation}
z_{\pm} \cdot z_{\pm} = {\rm Tr} \left( v_{\pm} {\cal C}^{-1} 
v_{\pm} {\cal C}^{-1} \right) 
= 0 \,, \qquad z_+ \cdot z_- = {\rm Tr} \left( v_{+} {\cal C}^{-1} v_{-} {\cal C}^{-1} 
\right) = 0\ .  \label{v-constr}
\end{equation}
Imposing them, one can check that the matrix $ww$ is zero as it 
should be according to (\ref{new-constr}).

Since we know that the HKC is flat, there should be coordinates 
$u_i;\,i=1,\ldots,4$, in which the hyperk\"{a}hler potential is
\begin{equation}
\chi = \sum_{i=1}^{4} u_{i} \bar{u}_{i}\ .\label{flat-pot}
\end{equation}
If we can find a gauge for $v_{\pm}$, satisfying \eqn{v-constr}, 
such that
\begin{equation}
w_{ij}=u_iu_j\ ,
\end{equation}
then \eqn{v-vdagger} reduces to \eqn{flat-pot} and we have proven that 
\eqn{cq} gives the correct result. 

This can be achieved by choosing the three complex $SU(2)$ gauge
conditions 
\begin{equation}
v_{+14} = v_{-13}\ , \qquad v_{+13} = v_{-14} = 0\ , \label{gchoice}
\end{equation}
and identifying
\begin{equation}
v_{+14} = v_{-13} = u_{1}\ , \qquad v_{+12} = -u_{3}\ , 
\qquad v_{-12} = u_{4}\ . \label{ident}
\end{equation}
Then we solve the moment map constraints (\ref{v-constr}) for 
$v_{+23}$, $v_{-24}$ and $v_{-23}$, and $w_{ij} = u_iu_j$ 
for $v_{+24}$. This gives
\begin{equation}
v_{+23} = \frac{u_{3}^{2}}{u_{1}}\ , \qquad 
v_{+24} = \frac{u_{3} u_{4}}{u_{1}} + u_{2}\ , \qquad       
v_{-23} = - \frac{u_{3} u_{4}}{u_{1}} + u_{2}\ , \qquad 
v_{-24} = - \frac{u_{4}^{2}}{u_{1}}\ . \label{solv}
\end{equation}
Using (\ref{invert}), (\ref{gchoice}), (\ref{ident}) and (\ref{solv}) it is straightforward to find the coordinates $z_{\pm}$ in terms of the flat coordinates $u_i$ for any explicit representation of the Clifford algebra. For example,
\begin{eqnarray}
&&z_{+} = \left(\, -2 u_{3} \,\, , \,\, -\frac{iu_{3}^{2}}{u_{1}} - iu_{1} \,\, , \,\, -\frac{u_{3}^{2}}{u_{1}} +u_{1} \,\, , \,\, \frac{u_{3}u_{4}}{u_{1}} + u_{2} \,\, , \,\, -\frac{iu_{3}u_{4}}{u_{1}} - iu_{2} \, \right)\ , 
\nonumber \\
&&z_{-} = \left(\,\,\,\,\, 2 u_{4} \,\, , \,\, \frac{iu_{3}u_{4}}{u_{1}} - iu_{2} \,\, , \,\, \frac{u_{3}u_{4}}{u_{1}} -u_{2} \, , \,\,\, -\frac{u_{4}^{2}}{u_{1}} + u_{1} \,\, , \,\,\,\,\,\,\,\, \frac{iu_{4}^{2}}{u_{1}} + iu_{1} \,\,\,\,\, \right)\ . \nonumber
\end{eqnarray}
One can easily check that these $z_{\pm}^{M}$'s satisfy (\ref{hol}) and give $\chi = \sum_{i=1}^{4} u_{i}\bar{u}_{i}$.

\subsection{$n=6$}

Recall that 
\begin{equation}
X(2) = \frac{SU(4)}{SU(2) \times U(2)}\ , \qquad 
Y(2) = \frac{SO(6)}{SO(2) \times SO(4)} \ ,
\end{equation}
and hence $X(2) = Y(2)$ and their HKC's must coincide. The hyperk\"{a}hler 
potential of the HKC of $X(n)$ was obtained in \cite{dWRV1} by 
a $U(1)$ hyperk\"{a}hler quotient of $\b{H}^{n+2}$. Denoting complex 
coordinates of $\b{H}^{4} = \b{C}^{8}$ by $w_{+}^{i}$ and $w_{-i}, 
\, i=1,...,4$, transforming in conjugate representations of $SU(4)$, 
we have
\begin{equation}
\chi(X) = 2 \sqrt{w_{+}^{i}\bar{w}_{+}^{i} w_{-j}\bar{w}_{-j}}\ , 
\qquad \mathrm{where} \qquad
w_{-}^{4}=1\, \,\,\, w_{+}^{4}=-w_{+}^{a}w_{-a}\, \,\,\, a=1,2,3\ . 
\label{X}
\end{equation}

To compare the HKC of $Y(2)$ with (\ref{X}), we follow the same strategy as 
for $n=5$, namely we make use of the local isomorphism $SU(4)=SO(6)$. 
To this end we define $U(1)$ gauge invariant (and traceless) matrix
\begin{equation}\label{g-inv-w}
w^i{}_j = w_{+}^{i}w_{-j}\, \qquad i,j=1,...,4\ ,
\end{equation}
and write (\ref{X}) in matrix notation as
\begin{equation}
\chi(X) = 2 \sqrt{\, {\rm Tr} (ww^{\dagger})}\, . \label{cx}
\end{equation}

On the $Y(2)$-side, the coordinates $z_{\pm}^M$, $M=1,...,6$ 
are in the vector representation of $SO(6)$, similarly to (\ref{so5}). 
As in the $n=5$ case, we rewrite the hyperk\"{a}hler potential (\ref{cq}) in $SU(4)$ notation. Then we make gauge choices and solve the holomorphic moment map constraints such that  \eqn{cq} coincides with (\ref{cx}).

We introduce a basis of selfdual and anti-selfdual four by four matrices $\eta,\bar\eta$ and write
\begin{equation}
v_{\pm\,ij}\equiv z_{\pm}^a {\eta}^a_{ij}+
i z_{\pm}^{a+3}{\bar \eta}^a_{ij}z\ ,
\end{equation}
where $a=1,2,3$ and $\eta^a$ and ${\bar \eta}^a$ are selfdual 
and anti-selfdual respectively, see {\it e.g.}, \cite{hooft}. We can 
raise and lower pairs of indices using $\epsilon^{ijkl}$,
\begin{equation}
*v_{\pm}^{ij}\equiv \ft12 \epsilon^{ijkl}v_{\pm\,kl}\ ,
\end{equation}
such that
\begin{equation}\label{invt}
z_{\pm}^a=\ft14 {\eta}^a_{ij}*v_{\pm}^{ij}\ ,\qquad
z_{\pm}^{a+3}=\ft{i}{4} {\bar \eta}^a_{ij}*v_{\pm}^{ij}\ .
\end{equation}
The $SU(2)$ gauge invariant variables can be put in the traceless matrix
\begin{equation}
v^i{}_j\equiv *v_+^{ik}\,v_{-kj}-*v^{ik}_-\,v_{+kj}\ ,
\end{equation}
and the hyperk\"ahler potential is then, up to a numerical factor,
\begin{equation}
\chi(Y(2)) = \sqrt{\, {\rm Tr}(vv^{\dagger})}\ .
\end{equation}
The holomorphic moment map constraints (\ref{hol}) take the form
\begin{equation}
{\rm Tr}(v_{+} *v_{+}) = 0 \qquad {\rm Tr}(v_{-} *v_{-}) = 0 \qquad {\rm Tr}(v_{+} *v_{-}) = 0. \label{muy}
\end{equation}

We must fix a gauge in which the matrices $v^i{}_j$ and $w^i{}_j$ of 
\eqn{g-inv-w} coincide. This can be done by setting
\begin{equation}
v_{+13} = v_{-12} = 0\ , \qquad v_{-13} = -\frac{1}{2}\ ,
\end{equation}
and identifying
\begin{equation}
v_{+12} = w_{+}^{1}\ , \qquad v_{+23} = -w_{+}^{3}\ , 
\qquad v_{-14} = \frac{1}{2} w_{-}^{3}\ .
\end{equation}
We then solve (\ref{muy}) in such a way that the matrices $w$ and $v$ 
become equal. The answer is
\begin{eqnarray}
&&v_{+14} = -w_{+}^{1} w_{-}^{2}\ , \qquad v_{+24} = w_{+}^{1} w_{-}^{2} 
+ w_{+}^{3} w_{-}^{3}\ , \qquad v_{+34} = -w_{-}^{2} w_{+}^{3}\ , \nonumber\\
&&v_{-23} = -\ft12 \frac{w_{+}^{2}}{w_{+}^{1}}\ , \qquad v_{-24} =-\ft12 
\frac{w_{-}^{3} w_{+}^{2}}{w_{+}^{1}}\ , \qquad v_{-34} = -\ft12 w_{-}^{2} 
- \ft12 \frac{w_{+}^{2} w_{-}^{2}}{w_{+}^{1}}\ . 
\end{eqnarray}
Using (\ref{invt}) it is straightforward to find $z_{\pm}^M$, $M=1,...,6$ 
in terms of $w_{\pm}^N$ $N=1,...,4$.

\section{The Dual Description of the HKC}
\setcounter{equation}{0}

As is well known, whenever there is an abelian isometry group one can find a dual description in terms of $N=2$ tensor multiplets. In this section, we consider the noncompact case, which is the one relevant for the low energy limit of string compactifications \cite{CFG}. The isometry group of $\tilde Y(n)$ is $SO(n,4)$ (see (\ref{S})). Since the coordinates $z_{+}^M, z_{-}^M, ~ M=1,...,n+4$ transform in the vector representation of this group, from (\ref{cq}) it follows that the hyperk\"{a}hler potential of the HKC of $\tilde{Y}(n)$ is also invariant, as are the two-forms (\ref{twoforms}).

To find the generators of a maximal abelian subgroup of $SO(n,4)$ note that any two matrices $T,T'$ of the form
\begin{equation}
T = \left( \begin{array}{ccc} 0 & 0 & a \\ 0 & 0 & a \\ a^t & -a^t & 0 \end{array} \right)\ , \label{T}
\end{equation}
where $a$ is an arbitrary $(n+2)$-dimensional row vector, commute with each other for any $a,a'$, and are nilpotent of third order. Such $T$'s are generators in the Lie algebra of the pseudo-orthogonal group that preserves   the metric $\eta=diag(-+\hat\eta)$, for any $\hat\eta$. If, for example, we arrange the coordinates as follows: $(1,5,2,3,...,n+4)$, then the flat metric on $\b{H}^{n,4}$ has this form, and we can write the generators of an $n+2$ dimensional abelian subgroup of $SO(n,4)$ as    
\begin{equation}
(T_I)_M{}^N = (\delta_M^1 + \delta_M^5) \delta_I^N \sigma_I + \delta_{MI} (\delta_1^N - \delta_5^N)\ , \qquad I \neq 1,5 \ , \label{Gen}
\end{equation}
where $\sigma_I$ is defined by $\eta_{MN} = \delta_{MN} \sigma_M$. It is easy to check that
\begin{equation}
(T_I)_M{}^L \eta_{LN} + (T_I)_N{}^L \eta_{LM} = 0~.
\end{equation}

We dualize with respect to the subgroup $U(1)^{n+2}$ generated by the set $\{T_I\}$ by starting with the flat space Lagrangian for $\b{H}^{n,4}$, gauging both the $n+2$ dimensional abelian symmetry found above as well as  the $SU(2)$ of Section 2, and constraining the $U(1)$ gauge fields with $N=1$ tensor multiplet Lagrange multipliers $G_I$. We also have to solve the holomorphic moment map constraints for the action of the $SU(2)$ (\ref{hol}) and those for the action of the $U(1)^{n+2}$:
\begin{equation}
z_{+M} z_{-N} \eta^{MK} (T_I)_K{}^N = v_I ~,  \label{v}
\end{equation}
where the $v_I$ are $N=1$ chiral superfields that together with $G_I$ make up an $N=2$ tensor multiplet. Finally, we gauge fix both $SU(2)$ and $U(1)^{n+2}$. The dual description is given by a Lagrangian which is a function of $v, \bar{v}$ and the real coordinates $G$. 

Let us see that all of the above requirements can be fulfilled consistently. 
Using (\ref{Gen}), (\ref{v}) becomes
\begin{equation}
(z_{+5} - z_{+1}) z_{-I} \sigma_I + (z_{-1} - z_{-5}) z_{+I} \sigma_I = v_I \, . \label{holu1}
\end{equation}
The $SU(2)$ action on each pair $(z_+^M,z_-^M)$ (\ref{var}) is fixed 
completely by setting
\begin{equation}
z_{+1} = z_{-5} \equiv \phi \, , \qquad z_{-1} = z_{+5} = 0 \, . \label{gauge}
\end{equation}
The $U(1)^{n+2}$ transformation rules on $\b{H}^{n,4}$, $\delta z_{\pm M} = \lambda_I (T_I)_M{}^N z_{\pm N}$, or explicitly 
\begin{equation}
\delta z_{\pm 1} = \lambda_I z_{\pm I} \sigma_I ~, \qquad \delta z_{\pm 5} = \lambda_I z_{\pm I} \sigma_I ~ , \qquad \delta z_{\pm I} = \lambda_I (z_{\pm 1} - z_{\pm 5}) \, ,
\end{equation}
acquire, on the HKC of $\tilde Y$, compensating $SU(2)$ transformations in order to preserve the gauge (\ref{gauge}), similar to (\ref{comp-su2}).
They  become
\begin{eqnarray}
&&\delta z_{+1} = \delta z_{-5} = \delta \phi = \frac{1}{2} \sigma_I \lambda_I (z_{+I} + z_{-I})~, \nonumber \\
&&\delta z_{+I} = \lambda_I \phi - \frac{1}{2} \sigma_J \frac{z_{+I} \lambda_J(z_{+J} - z_{-J})}{\phi} - \sigma_J \frac{z_{-I} \lambda_J z_{+J}}{\phi}~, \label{csu1} \\
&&\delta z_{-I} = - \lambda_I \phi - \frac{1}{2} \sigma_J \frac{z_{-I} \lambda_J(z_{+J} - z_{-J})}{\phi} - \sigma_J \frac{z_{+I} \lambda_J z_{-J}}{\phi}~. \nonumber
\end{eqnarray}
The holomorphic $SU(2)$ moment map constraints (\ref{hol}) reduce to
\begin{equation}
\phi^2 = z_{+I} z_{+J} \eta^{IJ} \, , \qquad \phi^2 = - z_{-I} z_{-J} \eta^{IJ} \, , \qquad 0 = z_{+I} z_{-J} \eta^{IJ} \, . \label{rc}
\end{equation}
We choose a $U(1)^{n+2}$ gauge
\begin{equation}
z_{+I} = 1 \qquad {\rm for~all} ~I \neq 1,5 ~ . \label{gauge1}
\end{equation}
Then (\ref{rc}) become
\begin{equation}
\phi^2 = n-4~ , \qquad \phi^2 = - z_{-I}^2 \sigma_I ~, \qquad z_{-I} \sigma_I = 0~,  \label{s}
\end{equation}
which we can solve for $\phi$ and two of the $z_{-I}$'s. The $U(1)^{n+2}$ holomorphic moment map (\ref{holu1}) after substituting (\ref{gauge}), (\ref{gauge1}) and the solution for $\phi$ is
\begin{equation}
- \sqrt{n-4} \, (z_{-I} + 1) \, \sigma_I = v_I \, . \label{zv}
\end{equation}
As a consequence of (\ref{s}), there are $n$ independent $v$'s.

The Lagrangian of the dual theory is
\begin{equation}
\sum_{M,N=1}^{n+4} \sigma_N \left( e^{ i \sum_{I} T_I V_I} \right)_{MN} \left( \bar{z}_{+}^M \,\, \bar{z}_{-}^M \right) 
\Big( \begin{array}{cc} {\rm e}^V \end{array} \Big)
\left( \begin{array}{c} 
z_{+}^N \\ z_{-}^N\end{array} \right) - \sum_{M \neq 1,5} \sigma_M G_M V_M ~ , \label{D}
\end{equation}
after we eliminate all gauge fields in (\ref{D}) via their equations of motion and impose the conditions (\ref{gauge}), (\ref{gauge1}) as well as the solutions to (\ref{s}), (\ref{zv}). Thus the dual description is in terms of $n$ $N=1$ chiral superfields $v$ and $n+2$ $N=1$ tensor multiplets $G$.

Unfortunately we haven't been able to solve for all gauge fields ($V$ and $V_I$) explicitly; and thus we have not been able to construct the dual Lagrangian. However the analysis that we made above shows that there is a consistent solution of all requirements (constraints and allowed gauge choices) for the existence of a dual description. In $N=2$ language, $n$ of the $G$'s combine with the $n$ $v$'s into $n$ $N=2$ tensor multiplets and the remaining two $G$'s combine into a double-tensor multiplet. This is consistent with the expectations from type IIB compactifications \cite{BGHL}. There the double-tensor multiplet appears naturally: it contains the dilaton, axion and the RR and NS two-forms.

Though we cannot solve for all gauge fields simultaneously, we can easily solve for either the $SU(2)$ prepotential $V$ or all $V_I$'s that gauge the $U(1)^{n+2}$. Eliminating the first gives the same answer for the dual Lagrangian as (\ref{cq}), but with $U(1)^{n+2}$ gauged as above. On the other hand, integrating out the $U(1)^{n+2}$ gauge fields gives us a dual description, although the result is still a function of the $SU(2)$ gauge fields, which for convenience we now write as $U = {\rm e}^V$. Using the nilpotency properties of the generators (and arranging the indices as $1,5,2,\dots ,n+4$), 
\begin{eqnarray}
{\rm e}^{i \sum_{I \neq 1,5} T_I V_I} &=& 1_{(n+4)\times (n+4)} + i \left( \begin{array}{ccccc} 0 & 0 & \sigma_2 V_2 & \sigma_3 V_3 & .. \\ 0 & 0 & \sigma_2 V_2 & \sigma_3 V_3 & .. \\ V_2 & -V_2 & 0 & 0 & .. \\ V_3 & -V_3 & 0 & 0 & .. \\ : & : & : & : & : \end{array} \right) + \nonumber \\
&-& \frac{1}{2}\left( \begin{array}{ccccc} \sum_{I \neq 1,5} \sigma_I V_{I}^2 & - \sum_{I \neq 1,5} \sigma_I V_{I}^2 & 0 & 0 & .. \\ \sum_{I \neq 1,5} \sigma_I V_{I}^2 & - \sum_{I \neq 1,5} \sigma_I V_{I}^2 & 0 & 0 & .. \\ 0 & 0 & 0 & 0 & .. \\ 0 & 0 & 0 & 0 & .. \\ : & : & : & : & : \end{array} \right)\ ,
\end{eqnarray}
we can rewrite (\ref{D}) as
\begin{equation}
\bar{Z}^{\alpha} M_{\alpha \beta} \sigma_{\beta} Z^{\beta} - \sigma_I G_I V_I
\ , \qquad \alpha, \beta = 1,...,2(n+4) \, .
\end{equation}
We have denoted $\bar{Z} = (\bar{z}_+^1, \, \bar{z}_-^1, \, \bar{z}_+^5, \, \bar{z}_-^5, \, \bar{z}_+^2, \, \bar{z}_-^2, \, ... \, , \, \bar{z}_+^{n+4}, \, \bar{z}_-^{n+4})$ and
\begin{equation}
M = \left( \begin{array}{ccccc} 
(1 - \frac{1}{2} \sum_{I} \sigma_I V_{I}^2) U & ( \frac{1}{2} \sum_{I} \sigma_I V_{I}^2) U & i \sigma_2 V_2 U & i \sigma_3 V_3 U & .. \\ 
( -\frac{1}{2} \sum_{I} \sigma_I V_{I}^2) U & (1 + \frac{1}{2} \sum_{I} \sigma_I V_{I}^2) U & i \sigma_2 V_2 U & i \sigma_3 V_3 U & .. \\ 
i V_2 U & -i V_2 U & U & 0 & .. \\ i V_3 U & -i V_3 U & 0 & U & .. \\ 
: & : & : & : & : \end{array} \right)~,
\end{equation}
where $M$ is a $(n+4)\times(n+4)$ matrix of $2\times2$ matrices. The solution for the $U(1)^{n+2}$ gauge fields is
\begin{equation}
V_I  = \frac{G_I - i \left( \bar{z}_{+}^1 + \bar{z}_{+}^5 \, , \, \bar{z}_{-}^1 + \bar{z}_{-}^5 \right) U 
 \left( \begin{array}{c} 
z_{+}^I \\ z_{-}^I \end{array} \right) + i \left( \bar{z}_{+}^I \, , \, \bar{z}_{-}^I \right) U \left( \begin{array}{c} 
z_{+}^1 + z_{+}^5 \\ z_{-}^1 + z_{-}^5 \end{array} \right)}{\left( \bar{z}_{+}^1 - \bar{z}_{+}^5 \, , \, \bar{z}_{-}^1 - \bar{z}_{-}^5 \right) U \left( \begin{array}{c} 
z_{+}^1 - z_{+}^5 \\ z_{-}^1 - z_{-}^5 \end{array} \right)}.
\end{equation}
Finally we note that if we dualize only $n+1$ instead of all $n+2$ generators of $U(1)^{n+2}$, we find $n+1$ $N=1$ tensor multiplets and $n+1$ $N=1$ chiral superfields, or equivalently, $n+1$ $N=2$ tensor multiplets.

\section{Half-hypermutliplets and $N=2$ Superspace}
\setcounter{equation}{0}

We close with a few comments on other representations of the HKC.
The half-hypermultiplet formulation that we have used so far is not easily described in $N=2$ superspace; however, a half-hypermultiplet has an alternative formulation as the double cover of an $A_1$ singularity, that is, a $U(1)$ hyperk\"ahler quotient of a full hypermultiplet doublet with vanishing Fayet-Iliopoulos terms.

Explicitly, consider a hypermultiplet doublet $z_\pm^{(1)},z_\pm^{(2)}$, with $z_+$ and $z_-$ in conjugate representations of a $U(1)$ gauge group with an $N=1$ gauge prepotential $V_0$ as well as the $SU(2)$ gauge group of the previous sections with an $N=1$ gauge prepotential $V$:
\begin{equation}
\chi={\rm e}^{V_0}\left( \bar{z}_{+}^{(1)} \,\, \bar{z}_{+}^{(2)} \right) 
\Big( \begin{array}{cc} {\rm e}^V \end{array} \Big)
\left( \begin{array}{c} 
z_{+}^{(1)} \\
z_{+}^{(2)} \end{array} \right)\  + {\rm e}^{-V_0} \left( {z}_{-}^{(1)} \,\, {z}_{-}^{(2)} \right) \Big( \begin{array}{cc} {\rm e}^{-V} \end{array} \Big)
\left( \begin{array}{c} \bar{z}_{-}^{(1)} \\ 
\bar{z}_{-}^{(2)} \end{array} \right)~.
\label{N2}
\end{equation}
This must be supplemented by the holomorphic moment map constraints; in particular, the holomorphic $U(1)$ moment map is \cite{dWRV1}
\begin{equation}
z_+^{(1)} z_-^{(1)} + z_+^{(2)} z_-^{(2)} = 0 \, . \label{choice}
\end{equation}
We shall see that integrating out $V_0$, imposing (\ref{choice}), and choosing a convenient gauge for the $U(1)$ symmetry gives us back the half-hypermultiplet formulation.

Integrating out the $U(1)$ gauge field $V_0$ gives \cite{dWRV1}
\begin{equation}
2 \, \sqrt{\left( \bar{z}_{+}^{(1)} \,\, \bar{z}_{+}^{(2)} \right) 
\Big( \begin{array}{cc} {\rm e}^V \end{array} \Big)
\left( \begin{array}{c} 
z_{+}^{(1)} \\ z_{+}^{(2)} \end{array} \right) \left( {z}_{-}^{(1)} \,\, {z}_{-}^{(2)} \right) \Big( \begin{array}{cc} {\rm e}^{-V} \end{array} \Big)
\left( \begin{array}{c} \bar{z}_{-}^{(1)} \\ \bar{z}_{-}^{(2)} \end{array} \right)} \label{3}
\end{equation}
We  can choose a $U(1)$ gauge
\begin{equation}
z_+^{(2)} = z_-^{(1)}\equiv z_-\, . \label{1}
\end{equation}
Then the holomorphic $U(1)$ moment map (\ref{choice}) implies
\begin{equation}
z_-^{(2)} = - z_+^{(1)}\equiv - z_+\, . \label{2}
\end{equation}
Using (\ref{1}) and (\ref{2}) we express everything in (\ref{3}) in terms of $z_\pm$:
\begin{equation}
2 \, \sqrt{\left( \bar{z}_{+} \,\, \bar{z}_{-} \right) 
\Big( \begin{array}{cc} {\rm e}^V \end{array} \Big)
\left( \begin{array}{c} 
z_{+} \\ z_{-} \end{array} \right) \left( {z}_{-} \,\, -{z}_{+}\right)
\Big( \begin{array}{cc} {\rm e}^{-V} \end{array} \Big)
\left( \begin{array}{c} \bar{z}_{-}\\ -\bar{z}_{+}\end{array} \right)} \, . 
\label{interm}
\end{equation}
The second factor in the square root can be rewritten as
\begin{eqnarray}
\left( {z}_{-} \,\, -{z}_{+} \right) \Big( \begin{array}{cc} {\rm e}^{-V} \end{array} \Big)
\left( \begin{array}{c} \bar{z}_{-} \\ -\bar{z}_{+} \end{array} \right) &=& \left( \bar{z}_{-} \,\, -\bar{z}_{+} \right) \Big( \begin{array}{cc} {\rm e}^{-V^T} \end{array} \Big)
\left( \begin{array}{c} z_{-} \\ -z_{+} \end{array} \right)  \nonumber \\ 
&=& \left( \bar{z}_{+} \,\, \bar{z}_{-} \right) \sigma_2  \Big( \begin{array}{cc} {\rm e}^{-V^T} \end{array} \Big) \sigma_2
\left( \begin{array}{c} z_{+} \\ z_{-} \end{array} \right)  \nonumber \\ 
&=& \left( \bar{z}_{+} \,\, \bar{z}_{-} \right) {\rm e}^{\Big( \begin{array}{cc}  -  \sigma_2 V^T \sigma_2 \end{array} \Big)}  \left( \begin{array}{c} z_{+} \\ z_{-} \end{array} \right)\, . \label{second}
\end{eqnarray}
Since $V$ is an $SU(2)$ gauge prepotential, we write it as $V=V^i \sigma_i \, ,i=1,2,3$, where $\sigma_i$ are the Pauli matrices, and
\begin{equation}
\sigma_2 \sigma_i^T \sigma_2 = -\sigma_i \, .
\end{equation}  
Hence (\ref{second}) is equal to the first factor under the square root in (\ref{interm}) and we confirm that (\ref{N2}) is the same as (\ref{hatchi}) (up to an insignificant factor of 2).

To complete the proof of the equivalence of (\ref{N2}) and (\ref{hatchi}), we must also compare the $SU(2)$ holomorphic moment maps. For (\ref{N2}) the $SU(2)$ moment map is \cite{HKLR}
\begin{equation}
\left( z_{-}^{(1)} \,\, z_{-}^{(2)} \right) T_{i} 
\left( \begin{array}{c} z_{+}^{(1)} \\ z_{+}^{(2)} \end{array} \right) = 0 \, ,
\end{equation}
where $T_i$ are the same generators as (\ref{gen}). Using (\ref{1}) and (\ref{2}) to express $z_{\pm}^{(1,2)}$ in terms of $z_{\pm}$, we get the same constraints as (\ref{hol}). 

Having this description in hand, we can now give the $N=2$ projective superspace formulation; we use the polar multiplet formulation (for a pedagogical summary as well as a definition of our notation, see the appendix B of \cite{dWRV1}; earlier references are contained therein and include \cite{projsup}). The HKC for $Y(n-4)$ has the projective superspace Lagrangian
\begin{equation}
L_{N=2} =\oint \frac{d\xi}{2\pi i} \sum_{i=1}^n {\rm e}^{{\bf V}_i}\bar\Upsilon_i {\rm e}^{\bf V}\cdot\Upsilon_i~,
\end{equation}
where each $\Upsilon_i$ is a doublet of polar multiplets, each ${\rm e}^{{\bf V}_i}$ is a tropical multiplet describing a $U(1)$ gauge prepotential (analogous to the $N=1$ gauge prepotential), and ${\rm e}^{\bf V}$ is a hermitian unimodular tropical multiplet describing the $SU(2)$ gauge prepotential. Formally, we may integrate out the latter and obtain:
\begin{equation}
L_{N=2} =\oint \frac{d\xi}{2\pi i} \, 2 \left(\det\left[\sum_{i=1}^n \Upsilon_i \otimes {\rm e}^{\bf V_i}\bar\Upsilon_i\right]\right)^{\frac12}~.
\end{equation}

The full hypermultiplet description can also be written as a quiver; indeed, the quiver for the HKC of $Y(1)$ was explored in quite some detail in the last section of \cite{LRvR}. The general quiver is a higher dimensional extension of (a cover of) the orbifold limit of the $D_4$ ALE space, and is shown in Figure 2:
\begin{figure}[htbp]
\centering
\leavevmode
\epsfxsize=2in
\begin{center}
\leavevmode
\ifpdf
\epsffile{quiver.pdf}  
\else
\epsffile{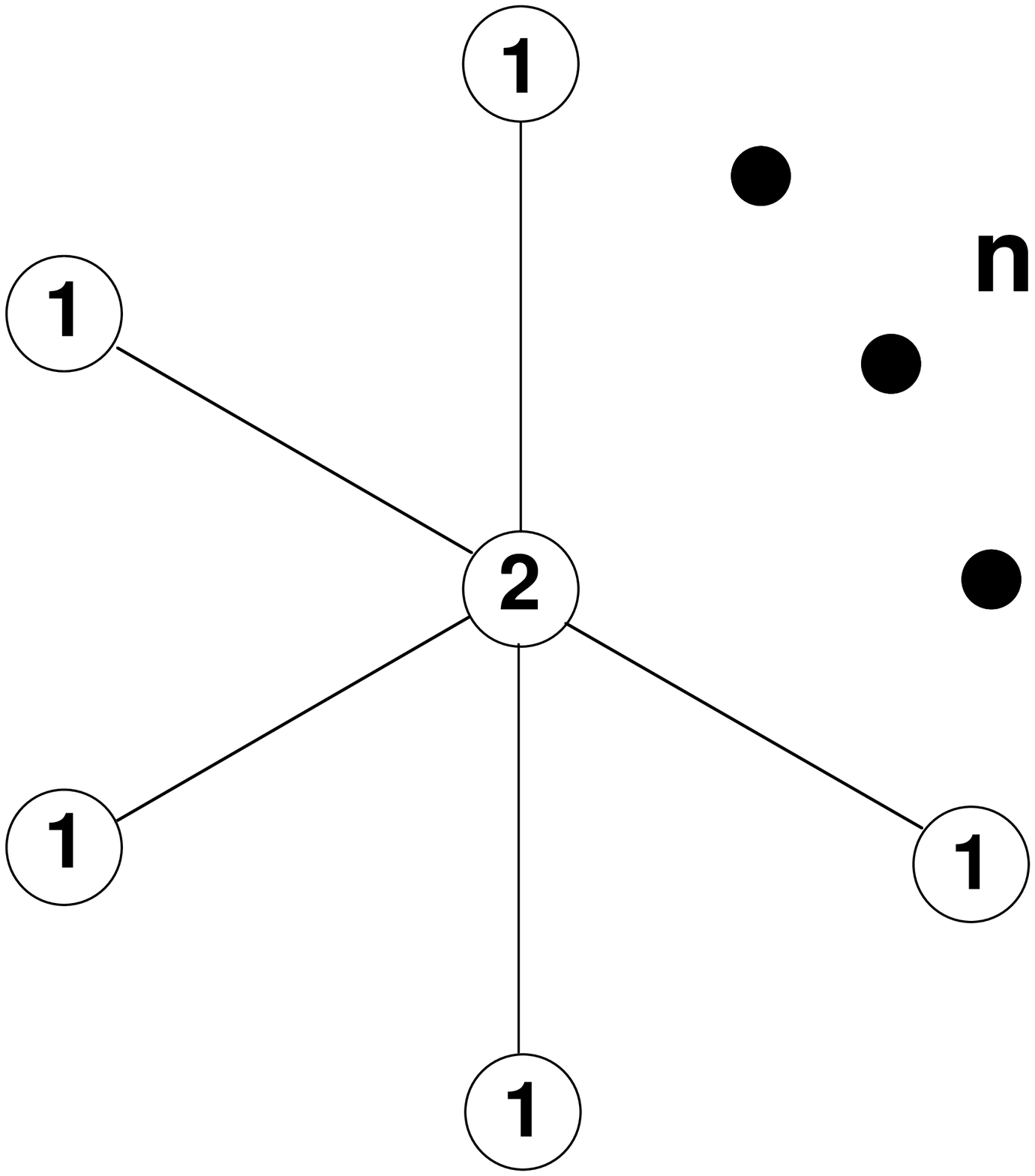}  
\fi
\end{center}
\vskip -.3cm
\caption{\footnotesize{The quiver describing the HKC of $Y(n-4)$. Each node with a label $k$ represents a factor $U(k)$ in the quotient group and each link is a hypermultiplet transforming in the bifundamental of the groups at the nodes it connects. An overall $U(1)$ acts trivially, and in our particular example, can conveniently be factored out by reducing the $U(2)$ at the central node to $SU(2)$.}}
\vskip .5cm
\end{figure}
The observation that the HKC of $Y(1)$ is flat implies that the specific quiver discussed in \cite{LRvR},  when the orbifold singularity is blown up (rather than removed by going to the cover) gives rise to a genuine 8-dimensional ALE manifold that has not been studied before. However, the higher dimensional analogs cannot be ALE, as the limit without any blowing up is not flat.
\vskip .5cm
\noindent{\bf \large Acknowledgements}

\noindent{SV thanks B. de Wit and M. Trigiante for stimulating discussions. 
MR thanks the NSF for partial support under grant NSF PHY-0098527.}
\eject

\end{document}